\documentclass[10pt,letterpaper,twocolumn,osajnl2,preprint,showpacs,superscriptaddress]{revtex4}
\usepackage[draft]{hyperref} 
\usepackage{mathrsfs}

\begin{document}

\title{Narrow-Band Biphoton Generation near Atomic Resonance}

\author{Shengwang Du}
\affiliation{Department of Physics, The Hong Kong University of Science and Technology, \\
Clear Water Bay, Kowloon, Hong Kong, China}
\email{shengwang.du@gmail.com}
\author{Jianming Wen}
\affiliation{Physics Department, University of Maryland, Baltimore County, Baltimore, Maryland 21250, USA}
\email{jianm1@umbc.edu}
\author{Morton H. Rubin}
\affiliation{Physics Department, University of Maryland, Baltimore County, Baltimore, Maryland 21250, USA}

\begin{abstract}
Generating nonclassical light offers a benchmark tool for the fundamental research and potential applications in quantum optics. Conventionally, it has become a standard technique to produce the nonclassical light through the nonlinear optical processes occurring in nonlinear crystals. In this review we describe using cold atomic-gas media to generate such nonclassical light, especially focusing on narrow-band biphoton generation. Compared with the standard procedure, the new biphoton source has such properties as long coherence time, long coherence length, high spectral brightness, and high conversion efficiency. In this paper we concentrate on the theoretical aspect of the entangled two-photon state produced from the four-wave mixing in a multilevel atomic ensemble. We show that both linear and nonlinear optical responses to the generated fields play an important role in determining the biphoton waveform and, consequently on the two-photon temporal correlation. There are two characteristic regimes determined by whether the linear or nonlinear coherence time is dominant. In addition, our model provides a clear physical picture that brings insight into understanding biphoton optics with this new source. We apply our model to recent work on generating narrow-band (and even subnatural linewidth) paired photons using the technique of electromagnetically induced transparency and slow-light effect in cold atoms, and find good agreements with experimental results.
\end{abstract}

\ocis{270.0270, 190.4410, 190.4380}

\maketitle

\section{Introduction}
Nonclassical light generation has attracted much attention over the last 40 years, partly because it not only provides a powerful probe for addressing fundamental issues of quantum theory such as complementarity, hidden variables, and other aspects central to the foundations of quantum mechanics \cite{bell}; but also holds promise for many potential applications to quantum information processing \cite{information}, quantum computation and communication \cite{Chuang}, quantum cryptography \cite{crytography}, quantum imaging \cite{Imaging}, quantum lithography \cite{Lithograhy1,Lithograhy2}, and quantum metrology \cite{Measurement}. In particular, entangled photon pairs have already been established as a standard research tool in the field of quantum optics. Traditionally, paired photons are produced from spontaneous parametric down conversion (SPDC) \cite{SPDC1,SPDC2}, in which a strong pump laser drives the atomic oscillators in a noncentrosymmetric crystal into a nonlinear regime, and then two down-converted beams are radiated by these oscillators. The two down-converted photons from such a nonlinear process usually have very broad bandwidth (typically in the THz range) and very short coherence time (typically around few picoseconds) so that their waveforms are not resolvable by existing single-photon detectors (which have resolution around the nanosecond range). In addition, the short coherence length of the SPDC biphotons, roughly about $\sim100$ $\mu$m, also limits their application in long-distance quantum communication. The optical properties of these wide-band biphotons have been well studied both experimentally and theoretically for more than two decades \cite{Rubin,Shih,klyshko}.

Narrow-bandwidth biphotons are ideal for a number of recently proposed protocols for long-distance quantum communication based on coherent interaction between single photons and atomic ensembles \cite{DLCZ, Lloyd}, which require efficiently absorbing biphotons and storing the entanglement. Other applications include the long-distance quantum state teleportation which requires long temporal coherence time of paired photons \cite{teleportation}. Using cavity-enhanced SPDC in nonlinear crystals, paired photons with a bandwidth of about 10 MHz and coherence time of about $50$ ns have been reported \cite{Cavity1,Cavity2,Cavity3,Cavity4}. However, the SPDC process is usually tuned to occur in the region of far-off-resonant atomic transitions, which limits the conversion efficiency and requires a large pump power. Moreover, in the far-off-resonant pumping the nonlinear parametric coupling coefficient can be treated as a constant and the information about the material internal transition structure is unresolvable from the two-photon correlation measurement.

The two most recently developed advanced technologies make the near- and on-resonance nonlinear processes possible. The first is laser cooling and trapping of neutral atoms \cite{Lasercooling}. Because cold atoms with a temperature of about 100 $\mu K$ have negligible Doppler broadening and very long lifetime, their atomic hyperfine structures can be resolved without need of a Doppler-free setup. The second is the electromagnetically induced transparency (EIT) \cite{EIT0,EIT1,EIT2} which not only can eliminate absorption on resonance but also enhance nonlinear interactions at low light level \cite{NLOHarris,NLOZhu,NLOBrajie,BrajeFWM}. Following the theoretical proposal in \cite{DLCZ} and early experiments by Lukin and Kimble groups \cite{Lukin, Kimble}, generation of nonclassical correlated photon pairs by exploring the ``writing-reading" procedure has attracted much research interest. However, the created paired photons are not maximally time-frequency entangled due to two separate operation processes. Using cold atomic-gas media and EIT-assisted four-wave mixing (FWM), Harris group at Stanford University was the first to generate time-frequency entangled narrow-band biphotons with cw driving fields in a double-$\Lambda$ system \cite{Balic, Kolchin}. By controlling the EIT at a sufficiently high optical depth (OD) around 50, biphotons with subnatural linewidth have been produced in a recent experiment, where the coherence time is up to about 1 $\mu$s in a two-dimensional magneto-optical trap (MOT) \cite{Subnatural}. Vuletic and his colleagues have also produced biphotons with a temporal coherence length of 100 ns by trapping cold atoms inside an optical cavity \cite{Vuletic}. Du \textit{et al.}, by driving a two-level system into the nonlinear region, have also created paired photons with correlation time of same order of magnitude \cite{Du}.

To describe the physics behind narrow-band biphoton generation from the EIT-based multilevel atomic system, two  different approaches have been presented in the literature. One approach is to use the Heisenberg-Langevin method by solving the coupled field operator equations in the Heisenberg picture, and the results agree well with the experimental data in the two-photon correlation measurement \cite{KolchinPRA,Ooi}. The advantage of this method is that it provides a sophisticated calculation of the effects from the Langevin noises, the medium absorption and optical gain on the biphoton wavepacket. Working in the interaction picture, the other approach developed by Wen \textit{et al.} \cite{Wen1,Wen2,Wen3,Wen4} focuses on the description of the two-photon state vector after the nonlinear medium using perturbation theory. This state vector theory not only properly describes the experimental results but also offers a clear physical picture on biphoton generation mechanism. It is the intent of this paper to review and generalize this latter approach by taking into account both linear loss and gain, and provide more insights on the roles of nonlinear interaction and linear response to the biphoton amplitude. Using this state vector picture, we show that the two-photon wave function is a convolution of the nonlinear and linear optical responses. According to this convolution, the two-photon temporal correlation is considered in two regimes, damped Rabi oscillation and group delay.

The paper is organized as the following. In Sec.~\ref{sec:BiphotonState}, using perturbation theory we lay out a general formulism of the two-photon state and biphoton wavepacket (or amplitude) generated from the FWM parametric process. In Sec.~\ref{sec:OpticalResponse}, taking a four-level double-$\Lambda$ atomic system as an example, we show that the linear and nonlinear optical responses to the generated fields play an important role in determining the two-photon amplitude. By looking at the nonlinear susceptibility, we illustrate the mechanism of biphoton generation in such a system. In Sec.~\ref{sec:SingleAtomRegime} we describe the damped Rabi oscillation regime where the biphoton amplitude is mainly characterized by the structure of the third-order nonlinear susceptibility. We show that the two-photon temporal correlation can exhibit damped and over-damped Rabi oscillaitons. In Sec.~\ref{sec:GroupDelayRegime}, we describe the group-delay regime where the biphoton wavepacket is mainly determined  by the phase matching due to the linear optical response. In Sec.~\ref{sec:Interference}, we examine the two-photon interference using biphotons generated from such a cold atomic-gas medium. We show that by manipulating the linear optical response, it is possible to switch the two-photon anti-bunching-like effect to bunching-like effect. Finally, we draw the conclusions and outlook in Sec.~\ref{sec:Conclusion}.

\section{\label{sec:BiphotonState}The Two-Photon State and Wave Function}

\begin{figure}
\includegraphics[width=8cm]{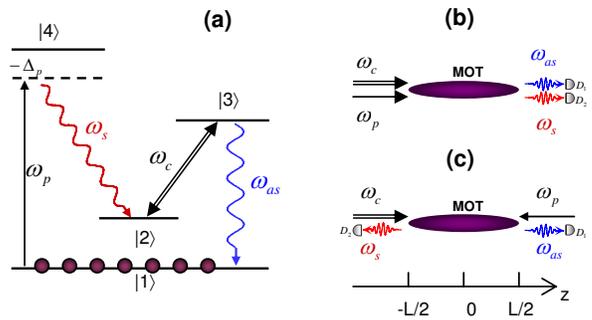}
\caption{\label{fig:FourLevelConfiguration}(color online) Biphoton generation via a four-level double-$\Lambda$ atomic system. (a) The level structure, where in the presence of a cw pump ($\omega_p$) and coupling ($\omega_c$) beams, paired Stokes ($\omega_s$) and anti-Stokes ($\omega_{as}$) photons are spontaneously created from the four-wave mixing processes in the low-gain regime. (b) The forward generation configuration. (c) The backward generation geometry. }
\end{figure}

A schematic of biphoton generation via a four-level double-$\Lambda$ atomic system is shown in Fig.~\ref{fig:FourLevelConfiguration}, where in the presence of a cw pump ($\omega_p$) and coupling ($\omega_c$) lasers, phase-matched, paired Stokes ($\omega_s$) and anti-Stokes ($\omega_{as}$) photons are spontaneously produced from the FWM process in the low-gain limit. As shown in Fig. \ref{fig:FourLevelConfiguration}(a), the strong coupling laser forms a standard three-level $\Lambda$ EIT scheme with the generated anti-Stokes field. Therefore, the role of the coupling laser here is that it not only assists the FWM nonlinear process but also creates a transparency window for the anti-Stokes photons by the slow-light effect. With the use of EIT, it is possible to generate the anti-Stokes photons near atomic resonance $|1\rangle\rightarrow|3\rangle$, which greatly enhances the efficiency of FWM process. The phase-matching condition allows forward and backward two generation configurations, as shown in Fig.~\ref{fig:FourLevelConfiguration}(b) and (c), respectively. In the backward generation configuration paired Stokes and anti-Stokes photons can propagate collinearly with the pump and coupling beams \cite{Balic,Subnatural}, and can also propagate in a right-angle geometry \cite{Kolchin}. These flexible generation setups depend on whether the relationship $\vec{k}_p+\vec{k}_c=0$ is well satisfied, where $\vec{k}_{p,c}$ are wave vectors of the input pump and coupling fields. One advantage for the backward generation geometry is that it allows us to easily separate the generated weak fields from the strong inputs.

In the following discussions, we assume that the cold atomic-gas medium consists of identical four-level atoms prepared in their ground state $|1\rangle$ [see Fig.~\ref{fig:FourLevelConfiguration}(a)]. The idealized atoms or molecules are confined within a long, thin cylindrical volume with  a length $L$ and atomic density is $N$. For simplicity, in this paper we will not take into account the Doppler broadening and polarization effects. In the two-photon limit, the quantum Langevin noise introduces unpaired photons which are not of interest here and so is ignored. In addition, we concentrate on the two-photon temporal correlation.

We start with the geometries shown in Fig.~\ref{fig:FourLevelConfiguration}(b) and (c), where Stokes-anti-Stokes photon pairs propagate along the $z$ axis. We denote the electric field as $E=\frac{1}{2}[E^{(+)}+E^{(-)}]=\frac{1}{2}[E^{(+)}+c.c.]$, where $E^{(+)}$ and $E^{(-)}$ are the positive- and negative-frequency parts. The input pump and coupling beams are assumed to be strong classical fields, so that
\begin{eqnarray}
E^{(+)}_p(z,t)&=&E_pe^{i[\pm k_p z-\omega_p t]},\nonumber \\
E^{(+)}_c(z,t)&=&E_ce^{i[k_c z-\omega_c t]},\label{eq:pcFields}
\end{eqnarray}
where $k_{p,c}$ are pump and coupling field wave numbers. The $\pm$ sign in Eq.~(\ref{eq:pcFields}) stands for the forward and backward propagation geometries for the pump laser. The single-transverse-mode Stokes and anti-Stokes fields are taken as quantized,
\begin{eqnarray}
\hat{E}^{(+)}_s(z,t)&=&\frac{1}{\sqrt{2\pi}}\int d\omega \sqrt{\frac{2 \hbar \omega}{c \varepsilon_0 A}}\hat{a}_s(\omega)e^{i[\pm k_s(\omega)z-\omega t]}, \nonumber \\
\hat{E}^{(+)}_{as}(z,t)&=&\frac{1}{\sqrt{2\pi}}\int d\omega \sqrt{\frac{2 \hbar \omega}{c \varepsilon_0 A}}\hat{a}_{as}(\omega)e^{i[k_{as}(\omega)z-\omega t]},
\label{eq:FieldOperators}
\end{eqnarray}
where $k_{s,as}$ are wave numbers of Stokes and anti-Stokes photons, and $A$ is the single-mode cross section area.. The annihilation operators $\hat{a}_s(\omega_s)$ and $\hat{a}_{as}(\omega_{as})$ satisfy the commutation relation
\begin{eqnarray}
[\hat{a}_s(\omega),\hat{a}^\dagger_s(\omega')]=[\hat{a}_{as}(\omega),\hat{a}^\dagger_{as}(\omega')]=
\delta(\omega-\omega'). \label{eq:FCommutation}
\end{eqnarray}
In the interaction picture, the effective interaction Hamiltonian for the four-wave mixing parametric process takes the form \cite{Wen1,Wen2,Wen3,Wen4}
\begin{eqnarray}
\hat{H}_I=\frac{\varepsilon_0A}{4}\int_{-L/2}^{L/2}dz\chi^{(3)}E^{(+)}_cE^{(+)}_p\hat{E}^{(-)}_{as}\hat{E}^{(-)}_s + h.c., \label{eq:Hamiltonian0}
\end{eqnarray}
where $h.c.$ means the Hermitian conjugate, and $\chi^{(3)}$ is the third-order nonlinear susceptibility to the Stokes (or anti-Stokes) field defined by the nonlinear polarizability $\hat{P}_{s,as}^{(3)(+)}=\varepsilon_0 \chi^{(3)}E_p^{(+)}E_c^{(+)}\hat{E}_{as,s}^{(-)}$ \cite{boyd}. With use of Eqs.~(\ref{eq:pcFields}) and (\ref{eq:FieldOperators}), after the $z$ integration we rewrite the Hamiltonian (\ref{eq:Hamiltonian0}) as
\begin{eqnarray}
\hat{H}_I&=&\frac{i\hbar L}{2\pi}\int d\omega_{as}d\omega_{s}\kappa(\omega_{as},\omega_s)\mathrm{sinc}\Big(\frac{\Delta k L}{2}\Big)\nonumber \\
&\times&\hat{a}^\dagger_{as}(\omega_{as})\hat{a}^\dagger_{s}(\omega_{s})e^{-i(\omega_c+\omega_p-\omega_{as}-\omega_s)t}
+h.c.,\label{eq:Hamiltonian}
\end{eqnarray}
where $\Delta k=k_{as}\pm k_s-(k_c\pm k_p)$ is the phase mismatching for the forward ($+$) and backward ($-$) configurations. When the pump and coupling fields are incident with an angle respect to the $z$ axis, the phase mismatching can be generalized as $\Delta k=(\vec{k}_{as}+\vec{k}_{s}-\vec{k}_{c}-\vec{k}_{p})\cdot\hat{z}$, where $\hat{z}$ is the unit vector along the $+z$ axis. The integral over the length $L$ gives a sinc function $\mathrm{sinc}(\frac{\Delta k L}{2})$, which determines the two-photon natural spectrum width. $\kappa(\omega_{as},\omega_s)=-i\frac{\sqrt{\varpi_{as}\varpi_s}}{2c}\chi^{(3)}(\omega_{as},\omega_s)E_pE_c$ is the nonlinear parametric coupling coefficient, where $\varpi_{as}$ and $\varpi_s$ are central frequencies of the anti-Stokes and Stokes fields, respectively.

Perturbation theory gives the photon state at the output surface(s) approximately as a linear superposition of  $|0\rangle+|\Psi\rangle$, where $|0\rangle$ is the vacuum state. The two-photon (biphoton) state $|\Psi\rangle$ can be expressed as
\begin{eqnarray}
|\Psi\rangle=-\frac{i}{\hbar}\int_{-\infty}^{+\infty}dt\hat{H}_I|0\rangle.\label{eq:BiphotonState0}
\end{eqnarray}
Because the vacuum is not detectable, from now on we will ignore it and only consider the two-photon part. Higher order terms with four, six, etc. photons are negligible for low-power continuous pumping. Using the Hamiltonian (\ref{eq:Hamiltonian}) in Eq.~(\ref{eq:BiphotonState0}), the time integral yields a $\delta$ function, $2\pi\delta(\omega_c+\omega_p-\omega_{as}-\omega_s)$, which expresses the energy conservation for the process. If $L$ is infinite, then the sinc function $L\mathrm{sinc}(\frac{\Delta kL}{2})$ in (\ref{eq:Hamiltonian}) becomes a $\delta$ function, $\delta(\Delta k)$. In this case, the conditions
\begin{eqnarray}
\omega_c+\omega_p-\omega_{as}-\omega_s=0,\nonumber\\
k_{as}\pm k_s-(k_c\pm k_p)=0 \label{eq:phasematching}
\end{eqnarray}
both hold and the phase matching is said to be perfect. The phase matching condition arises from the fact that the FWM process is a coherent process. Now the two-photon state (\ref{eq:BiphotonState0}) becomes
\begin{eqnarray}
|\Psi\rangle&=&L\int d\omega_{as} \kappa(\omega_{as},\omega_p+\omega_c-\omega_{as})\mathrm{sinc}\Big(\frac{\Delta k L}{2}\Big)\nonumber \\
&&\times\hat{a}^\dagger_{as}(\omega_{as})\hat{a}^\dagger_{s}(\omega_p+\omega_c-\omega_{as})|0\rangle.
 \label{eq:BiphotonState}
\end{eqnarray}
As seen from Eq.~(\ref{eq:BiphotonState}), the two-photon state is entangled in frequency and wave number, but is not entangled in polarization. In frequency space, the entanglement is the result of the frequency phase-matching condition, which implies that the detection of a photon at frequency $\omega_{as}$ requires the detection of the other photon at frequency $\omega_p+\omega_c-\omega_{as}$. The frequency correlation has interesting consequences for the temporal behavior of the pair \cite{Balic,Kolchin,Subnatural,KolchinPRA,Ooi,Wen1,Wen2,Wen3,Wen4,Precursor}, as we shall see. The state is also entangled with respect to the wave number since the sinc function $\mathrm{sinc}(\frac{\Delta k L}{2})$ cannot be factorized as a function of $k_{as}$ times a function of $k_s$. In the general noncollinear case, the wave-number entanglement has implication for the spatial correlation of photon pairs \cite{Wen1}. The polarization-entanglement generation has been discussed in \cite{Wen3,ThreeLevel}.

To study the optical properties of paired Stokes and anti-Stokes photons generated from such a four-level system, let us look at the simple experiments of two-photon joint-detection measurement as shown in Fig.~\ref{fig:FourLevelConfiguration}(b) and (c), where anti-Stokes photons go to detector D$_1$ and Stokes photons go to D$_2$. The two-photon temporal correlation is the quantity of primary interest in this paper. The field annihilation operators in time domain on the surfaces are expressed as
\begin{eqnarray}
\hat{a}_s(t)&=&\frac{1}{\sqrt{2\pi}}\int d\omega \hat{a}_s(\omega)e^{i[k_s(\omega)L/2-\omega t]}, \nonumber \\
\hat{a}_{as}(t)&=&\frac{1}{\sqrt{2\pi}}\int d\omega \hat{a}_{as}(\omega)e^{i[k_{as}(\omega)L/2-\omega t]}.
 \label{eq:OperatorInTime}
\end{eqnarray}
It can be shown that they obey the following commutation relation
\begin{eqnarray}
[\hat{a}_s(t),\hat{a}^\dagger_s(t')]=[\hat{a}_{as}(t),\hat{a}^\dagger_{as}(t')]=\delta(t-t').\label{eq:TCommutation}
\end{eqnarray}
The joint-detection probability is defined by Glauber's theory,
\begin{eqnarray}
G^{(2)}(t_{as},t_s)&=&\langle\Psi|\hat{a}_{s}^{\dagger}(t_s)\hat{a}_{as}^{\dagger}(t_{as})\hat{a}_{as}(t_{as})\hat{a}
_{s}(t_s)|\Psi\rangle \nonumber \\
&=&|\langle0|\hat{a}_{as}(t_{as})\hat{a}_{s}(t_s)|\Psi\rangle|^2 \nonumber \\
&=&|\Psi(t_{as},t_{s})|^2. \label{eq:Glauber}
\end{eqnarray}
In general, $\Psi(t_{as},t_{s})$ is referred to as the \textit{two-photon amplitude}, or \textit{biphoton wavepacket}.
With the help of Eqs.~(\ref{eq:BiphotonState}) and (\ref{eq:OperatorInTime}), the two-photon wave function (or amplitude) now is
\begin{eqnarray}
\Psi(t_{as},t_s)=\psi(\tau)e^{-i(\omega_c+\omega_p)t_s}, \label{eq:BiphotonWaveFunction}
\end{eqnarray}
where
\begin{eqnarray}
\psi(\tau)=\frac{L}{2\pi}\int d\omega_{as}\kappa(\omega_{as})\Phi(\omega_{as})e^{-i\omega_{as}\tau},
 \label{eq:BiphotonWaveFunction1}
\end{eqnarray}
with the relative time delay $\tau=t_{as}-t_{s}$ and $\kappa(\omega_{as})=\kappa(\omega_{as},\omega_{c}+\omega_{p}-\omega_{as})$. In Eq.~(\ref{eq:BiphotonWaveFunction1}), $\Phi(\omega_{as})$ is defined as the \textit{longitudinal detuning function},
\begin{eqnarray}
\Phi(\omega_{as})=\mathrm{sinc}\Big(\frac{\Delta kL}{2}\Big)e^{i(k_{as}+k_s)L/2}.\label{eq:detuningfunction}
\end{eqnarray}
From Eq.~(\ref{eq:BiphotonWaveFunction1}) we see that the biphoton wave function is determined by both the nonlinear coupling coefficient ($\kappa$) and the longitudinal detuning function ($\Phi$). In the time domain, $\psi(\tau)$ is a convolution ($\ast$) of $\tilde{\kappa}$ and $\tilde{\Phi}$
\begin{eqnarray}
\psi(\tau)=L[\tilde{\kappa}(\tau)\ast\tilde{\Phi}(\tau)]=L\int d\tau'\tilde{\kappa}(\tau')\tilde{\Phi}(\tau-\tau').
 \label{eq:convolution}
\end{eqnarray}
Here, $\tilde{\kappa}(\tau)$ is Fourier transform of the nonlinear coupling coefficient $\kappa(\omega_{as})$,
\begin{eqnarray}
\tilde{\kappa}(\tau)=\frac{1}{2\pi}\int{d}\omega_{as}\kappa(\omega_{as})e^{-i\omega_{as}\tau},
\label{eq:NonlinearResponse}
\end{eqnarray}
and similarly, $\tilde{\Phi}(\tau)$ is Fourier transform of the longitudinal detuning function $\Phi(\omega_{as})$,
\begin{eqnarray}
\tilde{\Phi}(\tau)=\frac{1}{2\pi}\int{d}\omega_{as}\Phi(\omega_{as})e^{-i\omega_{as}\tau}.\label{eq:NonlinearResponse}
\end{eqnarray}
The photon pair emission rate $R$ can be evaluated, by using Eqs.~(\ref{eq:BiphotonWaveFunction1}), as
\begin{eqnarray}
R=\frac{1}{2\pi}\int d\omega_{as}|\kappa(\omega_{as})\Phi(\omega_{as})|^2=\int d\tau|\psi(\tau)|^2.\label{eq:PairRate}
\end{eqnarray}

Now let us consider a photon counting experiment. Let the detectors be at equal distances from the source so that the relative time delay caused by the free-space propagation of electric fields are equal. Suppose the Stokes photon is detected by D$_2$ at time $t_{s}$, and after a controllable delay ($\tau$) detector D$_1$ is opened to detect the anti-Stokes photon. Assuming perfect detection efficiency, the averaged two-photon coincidence counting rate is determined by
\begin{eqnarray}
R_{cc}(\tau)&=&\lim_{T\rightarrow\infty}\frac{1}{T}\int^T_0dt_{s}\int^T_0dt_{as}\nonumber\\
&\times&G^{(2)}(t_{as},t_s)\Pi(t_{as}-t_{s};\tau,\tau+t_c)\nonumber\\
&=&\int d\tau'G^{(2)}(\tau')\Pi(\tau';\tau,\tau+t_c).
\label{eq:Rcc}
\end{eqnarray}
$\Pi(\tau';\tau,\tau+t_c)$ is the coincidence window function within the time bin width $t_c$: $\Pi=1$ for $\tau\leq\tau'\leq\tau+t_c$; otherwise, $\Pi=0$. When the two-photon coherence time is much larger than the detector time bin width $t_c$, Eq.~(\ref{eq:Rcc}) becomes $R_{cc}(\tau)\sim G^{(2)}(\tau)t_c$, which is a measure of the two-photon correlation function. This is the essence of narrow-band paired photons measurement discussed in this paper.

Equation (\ref{eq:BiphotonWaveFunction}) is the starting point for the discussions in the following sections. By extending wave numbers as complex variables, the effects of loss and gain inside the material can be properly included in the above theory, as we shall see. It is found that to be consistent with the Heisenberg-Langevin theory \cite{KolchinPRA,Ooi} in the low-gain limit, the argument in $\Phi$ should be replaced by $\Delta{k}=(\vec{k}_{as}+\vec{k}_{s}^*-\vec{k}_{c}-\vec{k}_{p})\cdot\hat{z}$, where $\vec{k}_{s}^*$ is the conjugate of $\vec{k}_{s}$.

Before proceeding the discussion let us make a short summary. In this section, we have laid out the general formulism of the two-photon state and wave function generated in a FWM medium pumped by two cw classical fields using perturbation theory. The two-photon state (\ref{eq:BiphotonState}) is a coherent superposition and it is entangled in frequency and wave numbers. The biphoton wave function, as shown in Eq.~(\ref{eq:convolution}), is a convolution of $\tilde{\kappa}$ and $\tilde{\Phi}$. Alternatively, Eq.~(\ref{eq:BiphotonWaveFunction1}) states that the two-photon spectrum is determined by both the nonlinear susceptibility and the phase matching. This implies two regimes for the temporal correlation measurement, as discussed in Sec.~\ref{sec:SingleAtomRegime} and Sec.~\ref{sec:GroupDelayRegime}. The methodology explored here is equivalent to the Heisenberg-Langevin theory \cite{KolchinPRA,Ooi} at the two-photon level. Moreover, this method offers a clear physical picture of the paired photons, which makes it easier to understand the two-photon interference experiment \cite{Interference1,Interference2,Interference3,Interference4}. Using the above theory new phenomena have been analyzed \cite{Subnatural,Precursor,WenFranson} which are not easy to be interpreted from the Heisenberg-Langevin theory. For example, the two-photon correlation at a high optical depth \cite{Subnatural}, the simulation by numerically solving the Heisenberg-Langevin coupled-operator equations shows a good agreement in the leading edge spike with the experiment data, but the nature of optical precursors is much easier to be understood in the present biphoton state picture \cite{Precursor}.

\section{\label{sec:OpticalResponse}Nonlinear and Linear Optical Responses}
As shown in Eq.~(\ref{eq:BiphotonWaveFunction}), the two-photon wavefunction is determined by both the nonlinear susceptibility and the longitudinal detuning function. Therefore, in this section we look at the linear and nonlinear optical responses to the generated fields. On the one hand, the importance of the nonlinear susceptibility is that it not only characterizes the strength of the nonlinear parametric process but also is the mechanism of biphoton generation. As a consequence, it determines the feature of the two-photon temporal correlation in one regime. On the other hand, the longitudinal detuning function (related to the linear susceptibilities) sets the natural spectral width of biphotons, and governs the pattern of the temporal correlation in another regime. The optical response of a four-level double-$\Lambda$ EIT system has been discussed in \cite{KolchinPRA,Ooi,Wen1}. Wen \textit{et al}. have further systematically studied the optical responses in a three-level double-$\Lambda$ scheme \cite{Wen3} and a two-level system \cite{Wen2}. Recently, the responses of a four-level inverted-Y system have been analyzed in \cite{Wen4}.

For simplicity, here let us still choose paired Stokes-anti-Stokes generation shown in Fig.~\ref{fig:FourLevelConfiguration} as an example to illustrate the physics behind FWM. Assuming that the intensity of the off-resonant pump laser is chosen so that the atomic population remains primarily in the ground level $|1\rangle$, the third-order nonlinear susceptibility for the generated anti-Stokes field is \cite{BrajeFWM, Wen1}
\begin{eqnarray}
\chi^{(3)}=\frac{N\mu_{13}\mu_{32}\mu_{24}\mu_{41}/(\varepsilon_0\hbar^3)}{(\Delta_{p}+i\gamma_{14})[|
\Omega_c|^2-4(\omega+i\gamma_{13})(\omega+i\gamma_{12})]}\nonumber \\
=\frac{-N\mu_{13}\mu_{32}\mu_{24}\mu_{41}/(\varepsilon_0\hbar^3)}{4(\Delta_{p}+i\gamma_{14})(\omega-\Omega_e/2+i
\gamma_{e})(\omega+\Omega_e/2+i\gamma_{e})},\label{eq:Chi3}
\end{eqnarray}
In Eq.~(\ref{eq:Chi3}), $\mu_{ij}$ are the electric dipole matrix elements, $\Omega_c=\mu_{23}E_c/\hbar$ is the coupling Rabi frequency, and $\gamma_{ij}$ are dephasing rates, respectively. $\Delta_p=\omega_p-\omega_{41}$ is the pump detuning from the atomic transition $|1\rangle\rightarrow|4\rangle$. $\omega=\omega_{as}-\omega_{31}$ is the detuning of the anti-Stokes photons from the transition $|1\rangle\rightarrow|3\rangle$, and we will take $\varpi_{as}=\omega_{31}$ as the anti-Stokes central frequency. $\Omega_e=\sqrt{|\Omega_c|^2-(\gamma_{13}-\gamma_{12})^2}$ is the effective coupling Rabi frequency. $\gamma_e=(\gamma_{12}+\gamma_{13})/2$ is the effective dephasing rate. The third-order nonlinear susceptibility has two resonances separated by $\Omega_e$ and each is associated with a linewidth of $2\gamma_e$. The two-photon coherence time is inversely proportional to $2\gamma_e$. More importantly, the two resonances here indicate two types of FWM processes. Alternatively, two types of paired Stokes and anti-Stokes photons can be generated from these two FWM processes in the spontaneous emission region. One FWM process happens when the central frequency of the anti-Stokes photons is $\varpi_{as}+\Omega_e/2$ and the central frequency of the correlated Stokes photons is $\omega_{42}-\Delta_p-\Omega_e/2$. The other FWM occurs as the anti-Stokes photons have the central frequency at $\varpi_{as}-\Omega_e/2$ while the paired Stokes photons have the frequency at $\omega_{42}+\Delta_p+\Omega_e/2$. This is the generation mechanism of biphotons produced from the FWM in atomic-gas media. Consequently, the interference between these two types of biphotons will appear in the two-photon temporal correlation, as shall be discussed in Sec.~\ref{sec:SingleAtomRegime}. The results obtained here agree well with the dressed-state picture \cite{Wen2,Wen3,Wen4}.

The linear susceptibilities at the Stokes and anti-Stokes frequencies are, respectively,
\begin{eqnarray}
\chi_s(\omega)=\frac{N|\mu_{24}|^2(\omega-i\gamma_{13})/(\varepsilon_0\hbar)}{|\Omega_c|^2-4(\omega-i\gamma_{13})(\omega
-i\gamma_{12})}\frac{|\Omega_p|^2}{\Delta_p^2+\gamma_{14}^2},\label{eq:chis}\\
\chi_{as}(\omega)=\frac{4N|\mu_{13}|^2(\omega+i\gamma_{12})/(\varepsilon_0\hbar)}{|\Omega_c|^2-4(\omega+i\gamma_{13})(
\omega+i\gamma_{12})},\label{eq:ChiLinear}
\end{eqnarray}
where $\Omega_p=\mu_{14}E_p/\hbar$ is the pump Rabi frequency. The complex wave numbers of Stokes and anti-Stokes photons are obtained from the relations $k_s=(\omega_s/c)\sqrt{1+\chi_s}$ and $k_{as}=(\omega_{as}/c)\sqrt{1+\chi_{as}}$, where the imaginary parts stand for the Raman gain and EIT loss, respectively. The linear susceptibility (\ref{eq:ChiLinear}) to the anti-Stokes field takes the form of standard EIT. As expected, the coupling beam $\Omega_c$ not only assists the propagation of the anti-Stokes photons by creating an EIT window, but also manipulates the group velocity using the slow-light effect. Another role played by EIT here is that it allows the nonlinear optics occurring near atomic resonance with small absorption and hence enhances the efficiency of the nonlinear interaction.

It is instructive to examine the linear responses under some simplifying approximation. Taking $|\Omega_p|\ll\Delta_p$, Eqs.~(\ref{eq:chis}) and (\ref{eq:ChiLinear}) give $\chi_s\simeq0$ and $k_{as}\simeq{k}_{as0}+\omega/V_g+i\alpha$ so that the wave-number mismatching is $\Delta k\simeq\omega/V_g+i\alpha$. Here $k_{as0}$ is the central wave number of the anti-Stokes field, $V_g$ is its group velocity, and $\alpha$, the imaginary part of the anti-Stokes wave number, characterizes the EIT loss, respectively. Now Eq.~(\ref{eq:detuningfunction}) can be approximated as
\begin{eqnarray}
\Phi(\omega)\simeq\mathrm{sinc}\Big(\frac{\omega L}{2V_g}+i\frac{\alpha L}{2}\Big)\exp\Big(i\frac{\omega L}{2 V_g}-\frac{\alpha L}{2}\Big),\label{eq:PhaseMatchingApp}
\end{eqnarray}
which sets a coherence time about $\tau_g\approx{L}/V_g$. In Eq.~(\ref{eq:PhaseMatchingApp}), $\alpha=2N\sigma_{13}\gamma_{12}\gamma_{13}/(|\Omega_c|^2+4\gamma_{12}\gamma_{13})$ where $\sigma_{13}=2\pi|\mu_{13}|^2/(\varepsilon_0\hbar\lambda_{13}\gamma_{13})$ is the on-resonance absorption cross section in the transition $|1\rangle\rightarrow|3\rangle$.

As discussed in this section, there are three important characteristic frequencies that determine the shape of the biphoton wave function. The first is the coupling effective Rabi frequency $\Omega_e$, which determines the two-resonance spectrum of the nonlinear susceptibility. The second is the linewidth $2\gamma_e$ of the two resonances in the nonlinear susceptibility. The third is the full-width-at-half-maximum (FWHM) phase-matched bandwidth determined by the sinc function, $\Delta\omega_{g}=2\pi\times0.88/\tau_g$, where $\tau_g$ is the anti-Stokes group delay time. In time domain, they correspond to the Rabi time $\tau_r=2\pi/\Omega_e$, nonlinear coherence time $\tau_e=1/(2\gamma_e)$ and group delay time $\tau_g$. The group delay time can be estimated from $\tau_g=L/V_g\simeq(2\gamma_{13}/|\Omega_c|^2)OD$ with the optical depth defined as $OD=N\sigma_{13}L$. The competition between $\tau_r$, $\tau_e$ and $\tau_g$ will determine which effect plays a dominant role in governing the feature of the two-photon correlation. Therefore, in the following we will discuss the two-photon joint-detection measurement in two regimes, damped Rabi oscillation and group delay.

\begin{figure*}
\includegraphics[width=16cm]{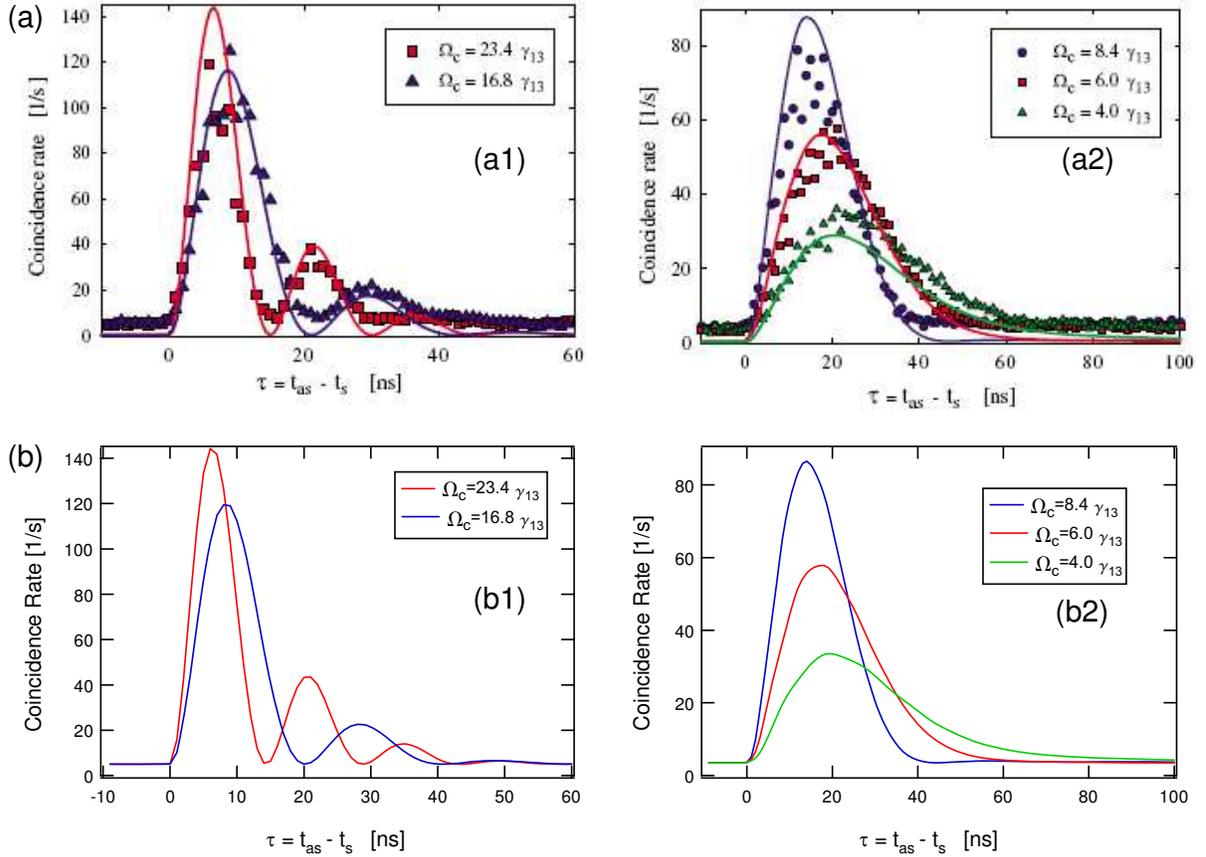}
\caption{\label{fig:RabiOscillation}(color online) Two-photon correlation function in the damped Rabi oscillation regime. (a) Experimental data and theoretical curves in (a1) and (a2) using the Heisenberg-Langevin theory from Ref.~\cite{Balic}. (b) Theoretical simulations using perturbation theory with same parameters from (a1) and (a2) and fitted with a vertical scaling factor and background accidental coincidence floor. The missing of other oscillation periods in (a2) and (b2) is due to the short dephasing time of about 33 ns. The experimental parameters used here are $\gamma_{13}=\gamma_{14}=2\pi\times3$ MHz, $\gamma_{12}=0.6\gamma_{13}$, $OD=$11, $\Delta_p=-7.5\gamma_{13}$, and $\Omega_p=0.8\gamma_{13}$.}
\end{figure*}

\section{\label{sec:SingleAtomRegime}Damped Rabi Oscillation Regime}
Let us first look at the damped Rabi oscillation regime, where optical properties of the two-photon amplitude (\ref{eq:BiphotonState}) are mainly determined by the nonlinear coupling coefficient. This regime requires that the effective coupling Rabi frequency $\Omega_e$ and linewidths $\gamma_e$ be smaller than the phase-matching bandwidth $\Delta\omega_g$ or $\tau_r>\tau_g$ and $1/(2\gamma_e)>\tau_g$, so that we can treat the longitudinal detuning function as $\Phi(\omega)\simeq1$ and $\tilde{\Phi}(\tau)\sim\delta(\tau)$. The biphoton spectral generation rate is proportional to $|\kappa L|^2\propto|\chi^{(3)} L|^2$. Hence both biphoton spectrum intensity and emission rate are proportional to $OD^2$. If $1/(2\gamma_e)>2\pi/\Omega_e$, the two-photon temporal correlation clearly exhibits a damped Rabi oscillation with oscillation period $2\pi/\Omega_e$; if $1/(2\gamma_e)<2\pi/\Omega_e$, the second and following oscillations are suppressed because of the short dephasing time, as experimentally demonstrated in \cite{Balic,Kolchin}.

Let us consider the case of $\Omega_c>|\gamma_{13}-\gamma_{12}|$ first, which implies a real effective coupling Rabi frequency $\Omega_e$. The two-photon wave function now is determined by Fourier transform of the nonlinear coupling coefficient (\ref{eq:NonlinearResponse}), which is
\begin{eqnarray}
\tilde{\kappa}(\tau)=B e^{-\gamma_e\tau}e^{-i\varpi_{as}\tau}\sin\Big(\frac{\Omega_e\tau}{2}\Big)\Theta(\tau),
\label{eq:f}
\end{eqnarray}
and $B=-i\frac{N\mu_{13}\mu_{32}\mu_{24}\mu_{41}\sqrt{\varpi_{as}\varpi_s}}{4c\varepsilon_0\hbar^3\Omega_e(\Delta_p+i
\gamma_{14})}$. The heaviside step function here is defined as $\Theta(\tau)=1$ for $\tau\geq0$, and $\Theta(\tau)=0$ for $\tau<0$. Combining with Eq.~(\ref{eq:BiphotonWaveFunction}), the biphoton wavepacket becomes
\begin{eqnarray}
\Psi(t_{as}=t_{s}+\tau,t_s)\simeq L \tilde{\kappa}(\tau)e^{-i(\omega_c+\omega_p)t_s}\nonumber\\
=i\frac{BL}{2}e^{-\gamma_e\tau}\big[e^{-i(\varpi_{as}+\Omega_e/2)t_{as}}e^{-i(\varpi_{s}-\Omega_e/2)t_{s}}\nonumber\\
-e^{-i(\varpi_{as}-\Omega_e/2)t_{as}}e^{-i(\varpi_{s}+\Omega_e/2)t_{s}}\big]\Theta(\tau),\label{eq:total}
\end{eqnarray}
where $\varpi_{as}+\varpi_{s}=\omega_p+\omega_c$ and $\varpi_{as}=\omega_{31}$ have been used. The physics of Eq.~(\ref{eq:total}) is understood as follows. Because the two-photon state is entangled, it cannot be factorized into a function of $t_{as}$ times a function of $t_s$. $|\Psi|^2$, depends only on the relative time delay $\tau$, which implies that the pair is randomly generated at any time. If the wave packets of the pump and the coupling beams are not plane waves, the term $e^{-i(\omega_c+\omega_p)t_{as}}$ would become a wave packet with a coherence length determined by the pump and coupling. Equation (\ref{eq:total}) shows also destructive interference between the two biphoton channels. The first term in the bracket on the RHS of Eq.~(\ref{eq:total}) represents the two-photon amplitude of paired anti-Stokes at $\varpi_{as}+\Omega_{e}/2$ and Stokes at $\varpi_s-\Omega_{e}/2$; while the second term is the two-photon amplitude of paired anti-Stokes at $\varpi_{as}-\Omega_{e}/2$ and Stokes at $\varpi_s+\Omega_{e}/2$.

To further see the interference, let us look at the two-photon Glauber correlation function,
\begin{eqnarray}
G^{(2)}(\tau)=\frac{1}{2}|B L|^2 e^{-2\gamma_e\tau}\big[1-\cos(\Omega_e\tau)\big]\Theta(\tau),\label{eq:RabiG2}
\end{eqnarray}
The physics of Eq.~(\ref{eq:RabiG2}) shows the beating (or interference) between two types of paired photons generated from the two FWM processes. The two-photon correlation (\ref{eq:RabiG2}) exhibits damped Rabi oscillations of period $2\pi/\Omega_e$. The damping rate is determined by the resonant linewidth $2\gamma_e$ in the doublet. At $\tau=0$, $G^{(2)}$ vanishes and as $\tau\rightarrow\infty$, $G^{(2)}$ also approaches to zero. This indicates photon anti-bunching-like effect [$G^{(2)}(0)\leq G^{(2)}(\tau)$]. In the counter-propagation geometry, if $\vec{k}_c+\vec{k}_p=0$, $G^{(2)}(\tau)$ becomes a symmetric distribution about $\tau=0$, because spontaneously emitted paired photons can trigger both detectors D$_1$ and D$_2$ as experimentally verified in \cite{Kolchin,Du}.

Figure~\ref{fig:RabiOscillation}(a) shows the first experimental demonstration of this type of biphoton source by Bali\'{c} \textit{et al.} in a cold $^{87}$Rb gas \cite{Balic}. The solid lines in Fig.~\ref{fig:RabiOscillation}(a1) and (a2) are theoretical curves obtained from the Heisenberg-Langevin theory. In Fig.~\ref{fig:RabiOscillation}(b1) and (b2), the curves are theoretical simulations using Eq.~(\ref{eq:RabiG2}) with the same experimental parameters in (a1) and (a2). Comparing with the Heisenberg-Langevin theory, the state vector picture also gives an agreement with the experimental data. From Eq.~(\ref{eq:RabiG2}), Fig.~\ref{fig:RabiOscillation}(a1) and (b1) correspond to the case of $\tau_e>\tau_r$ where the oscillations can be clearly observed before damping out; while Fig.~\ref{fig:RabiOscillation}(a2) and (b2) correspond to the case of $\tau_e<\tau_r$ where only the first oscillation is observable and other disappear due to the fast dephasing rate $\gamma_e=\frac{\gamma_{12}+\gamma_{13}}{2}$. Although in \cite{Balic} the lack of other oscillations in Fig.~\ref{fig:RabiOscillation}(a2) were interpreted due to the group delay, we believe that both cases shown in Fig.~\ref{fig:RabiOscillation}(a1)and (a2) are in the damped Rabi oscillation regime and the correlation time is well fitted with $\tau_e$ and $\tau_r$. This can be further verified by reducing the optical depth by a factor of 1/2, which is corresponding to a reduction of the group delay by a factor of 1/2. However the correlation time widthes in Fig.~\ref{fig:RabiOscillation}(a2) hold the same.

If $\Omega_c<|\gamma_{13}-\gamma_{12}|$, the effective coupling Rabi frequency becomes purely imaginary,  $\Omega_e=\pm{i}\beta_e$. The biphoton wave function becomes
\begin{eqnarray}
\psi(\tau)\simeq L \tilde{\kappa}(\tau)=B L e^{-\gamma_e\tau}e^{-i\varpi_{as}\tau}\sinh\Big(\frac{\beta_e\tau}{2}\Big)\Theta(\tau).
\label{eq:Overdamped1}
\end{eqnarray}
At the weak coupling limit, $\beta_e\rightarrow\gamma_{13}-\gamma_{12}$ and Eq.~(\ref{eq:Overdamped1}) takes the form
\begin{eqnarray}
\psi(\tau)\rightarrow\frac{BL}{2}e^{-i\varpi_{as}\tau}\big(e^{-\gamma_{12}\tau}-e^{-\gamma_{13}\tau}\big)\Theta(\tau).
\label{eq:Overdamped2}
\end{eqnarray}
This can be understood as follows. Since the coupling beam is very weak, the atomic-level structure is not changed due to the input power. Two types of biphotons are generated with the same central frequency but are associated with different linewidths, $\gamma_{12}$ and $\gamma_{13}$, respectively. The interference between these two types of biphotons gives a manifested exponential decay. In such a limit, the EIT disappears and the medium absorption decreases the biphoton generation efficiency. This case might be observable in the dilute atomic-gas medium with very low optical depth ($OD\ll1$).

\section{\label{sec:GroupDelayRegime}Group Delay Regime}
Suggested by Balic \textit{et al.}\cite{Balic, KolchinPRA} and demonstrated by Du \textit{et al.}\cite{Subnatural}, the group delay regime is defined as $\tau_g>\tau_r$ and the (EIT) slow light effect can be used to dynamically control the biphoton temporal correlation time. The group delay condition is equivalent to $\Omega_e>\Delta\omega_g$, i.e., the biphoton bandwidth is determined by the phase-matching spectrum. Therefore, in this section we treat the third-order nonlinear susceptibility as a constant over the phase-matching spectrum. As a consequence, the mechanism of double resonance of biphoton generation is washed out by the longer group-delay time. The two-photon correlation in this regime looks more like that with conventional SPDC photons. The coincidences tends to be a square-like pattern. The biphoton wave function can be approximated as $\psi(\tau)\simeq\kappa_0L\tilde{\Phi}(\tau)$ where $\kappa_0$ is the on-resonance nonlinear coupling constant. We find that in this regime, the EIT profile modifies the biphoton wave function through transparency window and slow light. When the EIT bandwidth ($\Delta\omega_{tr}\simeq|\Omega_c|^2/(2\gamma_{13}\sqrt{OD})$) is larger than the phase-matching bandwidth, the anti-Stokes loss can be ignored and the two-photon wave packet approaches a rectangular shape. However, when the EIT loss is significant, the biphoton wave packet follows an exponential decay.

When $\Delta\omega_{tr}>\Delta\omega_g$, we can ignore the loss in Eq.~(\ref{eq:PhaseMatchingApp}) and rewrite it as
\begin{eqnarray}
\Phi(\omega)\simeq\mathrm{sinc}\Big(\frac{\omega L}{2V_g}\Big)\exp\Big(i\frac{\omega L}{2V_g}\Big).
 \label{eq:PhaseMatchingGDR1}
\end{eqnarray}
Its Fourier transform gives the biphoton wave function
\begin{eqnarray}
\psi(\tau)\simeq\kappa_{0}L\tilde{\Phi}(\tau)=\kappa_{0}V_g \Pi(\tau;0,L/V_g)e^{-i\varpi_{as}\tau}.
\label{eq:SquareFunc}
\end{eqnarray}
The rectangular function $\Pi$, ranging from $\tau=0$ to $L/V_g$, shows that the anti-Stokes photon is always delayed with respect to its paired Stokes photon because of the slow light effect. It is clear that the two-photon correlation time is determined by the group delay $\tau_g=L/V_g$. The photon pair generation rate can be calculated from Eq.~(\ref{eq:PairRate}) as
\begin{eqnarray}
R=|\kappa_0|^2V_gL.\label{eq:PairRateGDR}
\end{eqnarray}
Thus even though the spectral generation rate, $\kappa_0L\Phi(\omega)$, scales as $OD^2$, the total rate of paired counts scales linearly as $OD$ because the bandwidth reduces linearly with optical depth \cite{Balic}. As explained by Rubin \textit{et al}. \cite{Rubin}, the rectangular waveform may be understood as follows: the paired photons are always produced from the same, but unknown, space point in the nonlinear medium. If emitted from the front surface, the anti-Stokes photon has no delay, and both photons arrive at detectors simultaneously. However, if emitted from the back surface, the anti-Stokes photon is delayed relative to the Stokes photon by $\tau_g$. For conventional SPDC photons, the rectangular-shaped biphoton wave packet has been well known and its sub-ps correlation time has been directly measured in \cite{Sergienko}.

Interestingly, we find that to observe the rectangular shape, the condition $\Delta\omega_{tr}>\Delta\omega_g$ sets a lower bound for the optical depth. Using $\Delta\omega_{tr}\simeq|\Omega_c|^2/(2\gamma_{13}\sqrt{OD})$ and $\Delta\omega_g\simeq2\pi/\tau_g\simeq\pi|\Omega_c|^2/(\gamma_{13}OD)$, we obtain $OD>4\pi^2$. It can be shown that such an optical depth is necessary to have the EIT delay-bandwidth product greater than two \cite{NLOHarris}.

\begin{figure}
\includegraphics[width=8cm]{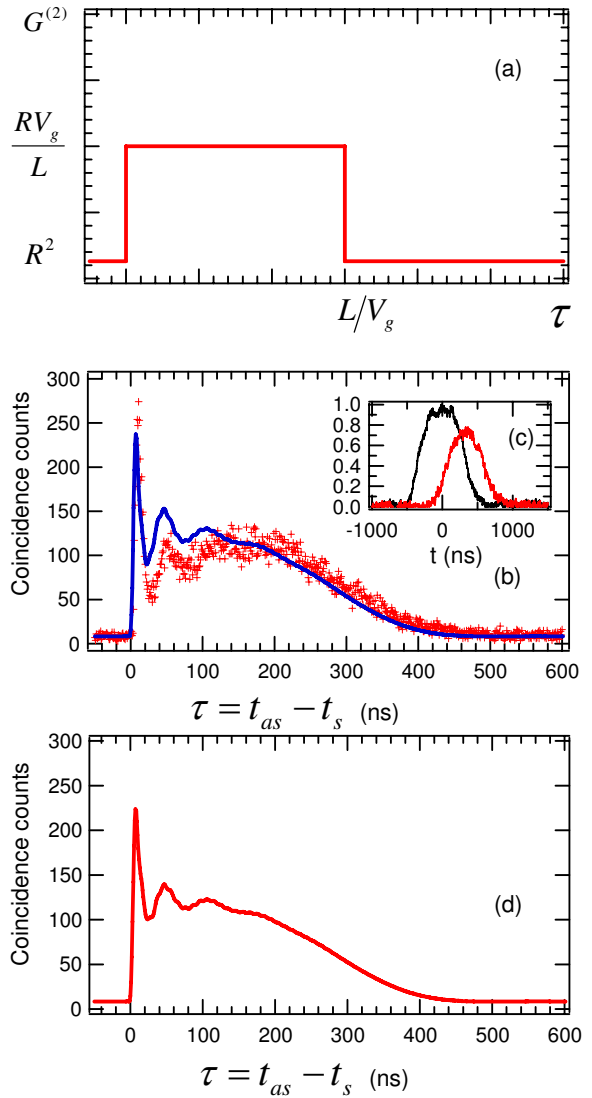}
\caption{\label{fig:GroupDelay}(Color online) Two-photon correlation function in the group delay regime. (a) Rectangular-shape correlation for an ideal spontaneous parametric down converter with a relative group delay $L/V_g$ and pair emission rate $R$. (b) Experimental data (red) and theoretical curve (solid blue line) obtained from the Heisenberg-Langevin theory in Ref.~\cite{Subnatural}. (c) Propagation delay of an anti-Stokes pulse (red) with respect to a reference pulse (black). Data are taken from Ref.~\cite{Subnatural}. (d) Theoretical simulation using the perturbation theory fitted with same experimental parameters in (b).}
\end{figure}

Figure~\ref{fig:GroupDelay}(a) shows an ideal rectangular-shape correlation with a group delay of $\tau_g=L/V_g$. In Fig.~\ref{fig:GroupDelay}(b) the experimental demonstration of a near-rectangular-shape correlation is obtained at the optical depth around 53 \cite{Subnatural}, where experimental data are denoted by ``+". The solid line is theoretical fitting using the Heisenberg-Langevin theory. As seen from Fig.~\ref{fig:GroupDelay}(b), the correlation time of about 300 ns is achieved and is consistent with the anti-Stokes pulse delay measurement shown in the inset (c), where the smaller curve represents the delayed signal and the higher one is the reference pulse. The experimental parameters used here are $\gamma_{13}=\gamma_{14}=2\pi\times3$ MHz, $\gamma_{12}=0.02\gamma_{13}$, $OD=53$, $\Delta_p=48.67\gamma_{13}$, $\Omega_p=1.16\gamma_{13}$, and $\Omega_c=4.20\gamma_{13}$, respectively. The EIT transparency width is estimated around $3.63$ MHz and the phase-matching spectral width about $2.93$ MHz. The theoretical simulation using the perturbation theory [Eq.~(\ref{eq:BiphotonWaveFunction1})], plotted in Fig.~\ref{fig:GroupDelay}(c), also fits very well with the reported experimental data. It is found that the exponential-decay behavior at the tail is due to the finite EIT loss, which alters the correlation function away from the ideal rectangular shape. In the experiment \cite{Subnatural}, Du \textit{et al.} have also experimentally verified that the total coincidence counts are linearly dependent on the optical depth.

The sharp peak shown in the leading edge of the two-photon coincidence counts in Fig.~\ref{fig:GroupDelay}(b) is the first observed Sommerfeld-Brillouin precursor \cite{Brillouin} at the biphoton level \cite{Subnatural,Precursor}. Mathematically, the integration in the two-photon amplitude (\ref{eq:BiphotonWaveFunction1}) is similar with the classical pulse propagation. In the stationary-phase approximation, starting from Eq.~(\ref{eq:BiphotonWaveFunction1}) one can show that the sharp peak is resulted from the interference of biphotons which are generated out of the EIT opacity window. The detailed analysis of the precursor generation at the two-photon level has been presented in \cite{Precursor} by using the theory laid out in Sec.~\ref{sec:BiphotonState}. Although by solving the coupled-operator equations the Heisenberg-Langevin theory predicts the correct pattern, the physics behind the precursor is not easy to be retrieved from this theory.

Suppose now the EIT loss is not negligible, i.e, $e^{-\alpha L}\ll1$. In such a case, the longitudinal detuning function (\ref{eq:PhaseMatchingApp}) can be approximated as
\begin{eqnarray}
\Phi(\omega)\simeq \frac{i}{\omega\tau_g+i\alpha L}.\label{eq:PhaseMatchingLoss}
\end{eqnarray}
The two-photon wave function becomes
\begin{eqnarray}
\psi(\tau)\simeq\kappa_0 V_ge^{-\alpha{V}_g\tau}e^{-i\varpi_{as}\tau}.\label{eq:DecayFunc}
\end{eqnarray}
The correlation function $|\Psi(\tau)|^2$ has an exponential decay time of $\tau_g/(2\alpha L)$. This means that the anti-Stokes photon detected at $\tau$ is generated at the depth of $V_g\tau$ such that the amplitude decreases by $e^{-\alpha V_g\tau}$.

\section{\label{sec:Interference}Two-Photon Interference: From Anti-Bunching to Bunching}
In Sec.~\ref{sec:SingleAtomRegime} and Sec.~\ref{sec:GroupDelayRegime}, we have discussed the two-photon temporal correlation in two regimes. In general, the two-photon wave function is a convolution of both the nonlinear and linear responses as shown in Eq.~(\ref{eq:convolution}). In both regimes, the correlation functions have non-zero values only at $\tau>0$. The observed anti-bunching-like effect in the damped Rabi oscillation regime is due to the destructive interference between two types of biphoton generation. In principle, if the detection system is frequency sensitive enough to resolve these two types of biphotons, the damped Rabi oscillations will disappear and the square-wave-like pattern manifested by exponential decay will be observed.

\begin{figure}
\includegraphics[width=8cm]{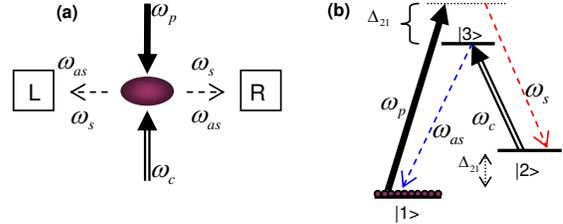}
\caption{\label{fig:Feyman}(color online) Biphoton generation in a three-level system. (a) Right-angle experimental configuration. (b) Three-level double-$\Lambda$ energy level diagram.}
\end{figure}

In this section, we are going to briefly show that it is possible to switch the two-photon coincidence counting rate from photon anti-bunching-like effect to bunching-like effect, by manipulating the linear optical responses to both Stokes and anti-Stokes fields \cite{ThreeLevel}. We consider a case that Stokes and anti-Stokes photons are degenerate at both spectrum and polarization so that both Stokes and anti-Stokes experience slow light propagation. As shown in Fig.~\ref{fig:Feyman}(a) and (b), we consider paired Stokes and anti-Stokes photons produced from a three-level double-$\Lambda$ scheme in a right angle configuration. In the counter-propagation geometry, frequency-phase matching allows both Stokes and anti-Stokes fields to reach both detectors. The two generated fields both may experience slow-light effect inside the medium due to the linear optical responses. The biphoton wave function now is expressed as
\begin{eqnarray}
\psi(\tau)\sim [\tilde{\kappa}(\tau)+\tilde{\kappa}(-\tau)]\ast\Pi(\tau;-L/V_g,L/V_g).\label{eq:convolution2}
\end{eqnarray}
Even though there is no temporal overlap between $\tilde{\kappa}(\tau)$ and $\tilde{\kappa}(-\tau)$, their convolution with $\Pi(\tau;-L/V_g,L/V_g)$ mixes them together in the time domain. By manipulating the group delay, the two-photon interference in Eq.~(\ref{eq:convolution2}) can be switched from the destructive to the constructive. Alternatively, the coincidence counting rate is changed from photon anti-bunching-like effect to bunching-like effect. This can be achieved by making the indistinguishability of their propagation pathways which gives rise to the biphoton interference. The interference comes from two indistinguishable paths for paired Stokes and anti-Stokes photons. In one possible propagation pathway, the pair is created at $z_1$ with Stokes photon going to the left and anti-Stokes photon to the right. In the other propagation pathway, the pair is created at $z_2=-z_1$ but with Stokes photon going to the right and anti-Stokes photon to the left. The interference between the two paths then modifies the two-photon correlation by further altering the group delay. The details of this two-photon interference are presented in \cite{ThreeLevel}. Recently, Wen \textit{et al.} has proposed a scheme to switch the two-photon interference from photon anti-bunching-like effect to bunching-like effect by manipulating the linear optical responses to the paired photons in a strongly driven two-level system \cite{WenFranson}.

\section{\label{sec:Conclusion}Conclusion and Outlook}
In conclusion, using perturbation theory, we have discussed the nature of the two-photon state and its waveform for narrow-band biphoton generation near atomic resonance via FWM parametric process using EIT and slow light. In general, the biphoton waveform is determined by a convolution of the nonlinear response (determined by the third-order nonlinear susceptibility) and the linear response (determined by the longitudinal detuning function). Therefore, the two-photon wave function can be analytically solved in two regimes, the damped Rabi oscillation and the group delay. In the damped Rabi oscillation regime, where the biphoton wave function is mainly determined by the third-order nonlinear susceptibility, the two-photon temporal correlation exhibits a damped Rabi oscillation because of the interference between two types of FWMs. In the group delay regime, where the two-photon amplitude is mainly determined by the phase-matching condition, the correlation approaches a rectangular shape manifested by an exponential decay. The theoretical simulations agree very well with recent experimental results \cite{Balic,Subnatural} and the Heisenberg-Langevin theory \cite{KolchinPRA,Ooi}. Although the discusses are based on a four-level double-$\Lambda$ EIT system, the analysis can be easily extended to three-level and two-level systems.

Before concluding, we note that a number of other interesting avenues are presently being explored that are closely related with this new type of biphoton source. By manipulating the linear optical responses to the generated fields,  Bell inequality can be tested by two-photon beatings using biphotons generated from a two-level system \cite{WenFranson}. It is interesting to note that the correspondence principle fails in describing paired-photon generation in a two-level system. As mentioned in \cite{Du,Wen2}, it is not generally valid to simply replace the coupled-field Maxwell's equations by coupled field-operator equations without changing the coefficients. But the underlying physics is not clear at this moment.

It will be interesting to explore quantum networks with these narrow-band paired photons. For example, the feasibility of an EIT-based quantum memory can be easily demonstrated using current biphoton source by looking at the two-photon interference, as suggested in \cite{wen2004}. It may be interesting to look at the squeezing properties after an EIT system if the input squeezed light is initially generated from the atomic-gas medium \cite{sue}. As proposed in \cite{Wen3,ThreeLevel} experimental demonstration of polarization-entangled narrow-band biphoton generation would be useful for quantum computation and communication.

Another problem is that it is not clear whether the maximally time-frequency entangled photon pairs can be generated in the room-temperature atomic-gas medium. Since all the experimental and theoretical work now are focused on cold atomic ensembles, it would be interesting to know if the Doppler broadening becomes dominant in the process, what kind of feature would be observed in the two-photon correlation measurement? By optimizing the system, can we still observe the damped Rabi oscillations?

Apart from temporal correlation, a possible application of these narrow-band biphotons may involve the transverse correlation \cite{Wen1}. It may be worth studying whether this new type of source is useful for quantum imaging and lithography.

\section{Acknowledgements}
S. Du acknowledges the start-up financial support from the Department of Physics at the Hong Kong University of Science and Technology. J.-M. Wen and M. H. Rubin were supported in part by the U.S. Army Research Office under MURI Grant W911NF-05-1-0197.

\end{document}